\documentstyle[12pt,aasms4,flushrt]{article}

\slugcomment{submitted to The Astrophysical Journal}

\lefthead{A. Gonz\'{a}lez}
\righthead{Tidal shear}

\begin{document}

\title{Large-scale tidal fields on primordial density peaks ?}

\author{Alejandro Gonz\'{a}lez}
\affil{Coord. de Astrof\'{i}sica, Instituto Nacional de 
       Astrof\'{i}sica, Optica y Electr\'{o}nica. \\
       A.P. 51 y 216, C.P. 72000, Tonantzintla, Puebla, M\'{e}xico.}




\begin{abstract}
We calculate the strength of the tidal field produced by the large-scale 
density field acting on primordial density perturbations in power law models. 
By analasing changes in the orientation of the deformation tensor, resulted
from smoothing the density field on different mass scales, we show  
that the large-scale tidal field can strongly affect the morphology and 
orientation of density peaks. The measure of the strength of the tidal field 
is performed as a function of the distance to the peak and of the spectral 
index. We detected evidence that two populations of perturbations seems to
coexist; one, with a misalignment between the main axes of their inertia 
and deformation tensors. This would lead to the angular momentum acquisition 
and morphological changes. For the second population, the perturbations are 
found nearly aligned in the direction of the tidal field, which would imprint 
them low angular momentum and which would allow an alignment of structures 
as those reported between clusters of galaxies in filaments, and between 
galaxies in clusters. Evidence is presented that the correlation between the 
orientation of perturbations and the large-scale density field could be a 
common property of Gaussian density fields with spectral indexes $n < 0$. We 
argue that alignment of structures can be used to probe the flatness of the 
spectrum on large scales but it cannot determine the exact value of the 
spectral index.

\end{abstract}


 

%

\section{Introduction}

At the structure formation, density perturbations
gravitationally interacted and were subject to the influence of tidal fields.
which are generated by the anisotropic distribution of neighbouring 
perturbations. The evolution and response of density peaks to the tidal 
field mainly depends on i) the shape of the perturbation, ii) the strength 
of the external tidal field and 
iii) the relative orientation between the shape and the field. Concerning 
shapes, Peacock and Heavens (1985), Bardeen et al. (1986) and  
Gonz\'{a}lez (1994) have confirmed the intrinsic triaxial morphology 
of peaks in Gaussian random fields. The tidal field, then, get coupled with the 
quadrupole momentum of the perturbation, exerting a torque and imprinting 
it angular momentum. The torque per unit mass at a distance $R$ from the 
perturbing body decreases as $R^{-3}$, but increases proportionally to the 
mass. If the power spectrum of perturbations is flat on large scales 
(Vogeley et al., 1992), then this mass increases as $R^{3}$ 
and therefore a divergence in the total torque could in principle appear. 
This would mean that the local properties of density fluctuations can be 
affected by the large-scale density field (see also Barnes, 1992). We study 
this possibility.

The orientation of incipient perturbations is also of great importance 
because it is closely related to the amount of angular momentum to be 
acquired. Density perturbation which are born with their major axes 
aligned with those of the tidal tensor experience a null or weak torque, 
and therefore would acquire a low amount of angular momentum. It is 
possible, in principle, that some of these perturbations preserve their 
initial orientations even 
after the non-linear evolution of the density field. Whether the initial 
number of these peaks is statisticaly significant or not is a subject which 
has not been studied so far. There are some observational and theoretical 
evidences which suggest that this number could be important in the past 
and is preserved today.

Interesting results in the above direction are the analysis of the relative 
orientation between the major axes of galaxies and that of their 
host cluster (e.g. Adams, Strom and Strom, 1986 ) and the orientation between 
clusters (West 1989, Plionis et al. 1992, Plionis 1994). For example, Lambas, 
Groth and Peebles (1988, LGP) studied the orientation of galaxies 
relative to the large-scale structures in which they are embedded (see also 
Djorgovski 1983, 1986). Not only an important tendency of the mayor axes of 
galaxies to be aligned with their host larger structures was detected, 
extending up to $\sim 4-5h^{-1}$Mpc, but also this effect was found  
morphological type dependent: alignments were detected for elliptical 
galaxies but not for spirals. 
In the explanation of the origin of the anisotropy in the orientation of 
galaxies, LGP discounted the action of the tidal field produced by 
neighbouring galaxies. They argue that if tidal fields were responsibles for 
the alignment of galaxies, then the same degree of alignment of spirals and 
ellipticals would be expected.  Or alternatively, a correlation would be 
expected between the orientation of a galaxy and its closest neighbour, 
independently of their morphological type: none of these tendencies was 
observed. Moreover, in the N-body simulations carried on by Barnes and 
Efstathiou (1987), they found that the alignment of the major axes of the 
formed objects is more prominent than is the alignment of angular momentum 
vectors of nearby objects. As pointed out by Barnes and Efstathiou (1987)  
``{\it The most striking effect seen in these tests is the tendency for 
nearby objects to point at each other. It seems likely that objects are 
born with these orientations; if they had been formed  with random major 
axes and sheared into line by tidal torques, we would expect a strong 
spin-vector versus separation-vector effect, which we do not find}''.
If tidal torques between neighbouring structures have not been a determining
factor in the generation of alignments, then how do these coherent patterns 
arise ? Furthermore, LGP have also 
suggested that {\it the orientation of ellipticals galaxies could reflects 
the primordial orientation of maxima of the density field. In such a case, 
the angular momentum vector of spiral galaxies need not have any connection 
with the orientation of elliptical galaxies}. The density peaks from which 
ellipticals form would be oriented in the direction imposed by the large-scale 
density perturbations to  which they belong. This idea has also been suggested 
several times by Bond (1986, 1987a,b) to explain the paralel orientation of cD 
galaxies with the cluster major axes. In a similar direction West (1994) 
proposed that cD's formed from special seeds and orientations.

It is precisely the primordial tidal field and the primordial alignment of 
density perturbations which we will investigate in this paper. In Sect. 2 
the influence of the far density field on the local properties of galactic 
and cluster scale peaks are studied, as a function of the distance and the 
spectral index of the density field; in Sect. 3 a measure of the strength 
of the tidal shear is carried on by considering two methods; in Sect 4 a 
discussion of the results and their possible consequences is presented.

\section{Is there any correlation between the large and 
         small-scale density field ? }

\subsection{Basic Formalism and Procedure}

Properties such as the amplitude, the shape, the orientation and the strength 
of the deformation tensor of density peaks are determined by the superposition
of density waves. One way of quantifying the influence coming from the 
large-scale field on galactic-scale peaks, consists in measuring changes 
of these properties as the density field is smoothed on increasing mass 
scales $M=M(R_{f})$. Where the only physical consequence of using a larger 
radius of filtering, $R_{f}$, is the generation of more massive perturbations 
of a larger characteristic radii. Once the primordial density field is 
generated within a $64^{3}$ cubic grid, and smoothed with a Gaussian filter 
function, the peaks mass grows as (e.g., Bardeen et. al. 1986)
\begin{equation}
      M(R_{f})=(2\pi)^{3/2}\bar{\rho}R_{f}^{3}=4.37\times 
      10^{12} R_{f}^{3}h^{-1}M_{\odot}.
\end{equation}
A radius of $R^{(1)}_{f}\approx 0.6 {\rm Mpc}\equiv 1$ 
corresponds with a galactic mass scale $M \approx 10^{12} M_{\odot}$. For
rich clusters, $R^{(1)}_{f}\approx 10 {\rm Mpc}\equiv 1$ yields 
$M\approx 10^{15} M_{\odot}$. 
Next, we identify the positions ${\bf x}_{i}$ of the density peaks.
Suppose now, that the density field is smoothed on a larger scale by using
$R^{(2)}_{f}=2 (\approx 1.3 {\rm Mpc})$. If changes in the properties of the
peaks at the positions ${\bf x}_{i}$ are observed, then they can be attributed 
to the ramaining superposition of density waves bounded by the spherical shell 
defined by the difference 
$\Delta R_{f}= R^{(2)}_{f}-R^{(1)}_{f}$. 
Intuitively, one expect that when $R^{(2)}_{f}\approx R^{(1)}_{f}$, the 
general properties of a given density peak do not change considerably. The 
more distant shells are specified by taking two consecutive filtering scales 
as $R^{(i+1)}_{f}-R^{(i)}_{f}$. The filtering $R_{f}^{(8)}$ represents the 
largest scale of $4.8$Mpc when we analize the effect on galactic-scale peaks, 
and $80$Mpc when we analize the effects on cluster-scale peaks. Throghout 
this paper we consider an Einstein-de Sitter $(\Omega=1)$ 
Universe.  

In the present analysis we restric ourselves to quantify the changes in i) the 
amplitud of the perturbations and ii) the changes in orientation and strength 
of the their deformation tensor. Thus, for $R^{(1)}_{f}$ and a given 
perturbation, we shall evaluate the angle $\theta$ subtended by the major axes 
associated with the unperturbed and the perturbed deformation tensor, this 
latter being the result of increasing the smoothing scale. If no changes
are observed, it would be indicative that the density waves of very long 
wavelength have not important effect on the galactic-scale peaks.
The deformation tensor is calculated from the Zel'dovich (1970) 
approximation as  
\begin{equation}
  {\bf D}_{jk}= \delta_jk + b(t)\frac{\partial v_{k}}{\partial q_{j}},
\end{equation}
where the values of the density and the velocity field at points
outside the grid vertexes are obtained by interpolation. The validity of the 
computed peculiar velocity is checked out through the continuity equation
(Peebles 1980). Finally, 
before going on into the results, we shall remark that the discussion
will be adressed to power law spectra, $P(k)\propto k^{n}$, with $n= 1, 0, 
-1, -2$. The spectrum for the Cold and Mixed Dark Matter models can be fitted 
through a power law with some appropriate index. For instance, the predicted 
shape of the CDM fluctuations 
spectrum corresponds with $P(k)\propto k^{-3}-k^{-2}$ on subgalactic and 
galactic scales, and  $P(k)\propto k^{-1}- k^{0}$ on the scale of rich 
clusters and superclusters (Bond and Efstathiou 1984; West 1994).

\subsection{Results}


Figure 1 is an example of the changes produced in the relative orientation 
of the deformation tensor for some of the highest peaks. Meanwhile, the 
sequence of histograms of Figure 2 shows the total number of peaks 
$N_{pk}$ with changes in 
$\cos\theta$ as a function of the mass scale. It is observed that the nearest 
shells produce the most important changes in the orientation 
of the deformation tensor as can be deduced from the changes produced 
in the form of the histograms. The influence on the peaks decreases 
with the distance of the shell of mass considered. When distant shells, and 
therefore the large-scale density field, are taken into account the form of the 
histograms remains unaltered for the steeper spectra, but not so for the flatter 
ones. To clarify this point, consider the form of the histograms for the cases
$n=1, 0$. The maximum of the distributions of peaks lies close to
$\cos\theta=1$, corresponding with small changes in the orientation of the
deformation tensor. For these spectral indexes, the number of peaks within the
interval $0 < \cos\theta < 0.9$, for which the orientation changes are
larger than $\sim 25^{o}$, includes approximately $60-70\%$ of the total 
sample for $R^{(2)}_{f}-R^{(1)}_{f}\approx 1.6$Mpc and decreases to 
$20-30\%$ for $R^{(3)}_{f}-R^{(2)}_{f}\approx 2.4$Mpc. This means that effects on 
density peaks are mainly due to the neighbouring distribution of mass to this scale.


For flat spectra, $n=-1, -2$, the situation is different. The form of the 
initial distributions for the nearest shell, is modified as one considers larger
filtering scales. The peak of the histograms becomes sharper near $\cos\theta=1$
only to scales larger than $\approx 3$Mpc. A large number of peaks present 
changes in the orientation of the deformation tensor even for shell distances of 
$\approx 3-5$Mpc. Such scales nearly matches the scale of a cluster of galaxies. 
Thus, a correlation between a cluster-scale perturbation and its central regions 
would be feasible. The numerical simulations by Faber (1982), Blumenthal et al. 
(1984) and White \& Frenk (1991) suggest that a power spectrum index of $n\approx 
-2$ is consistent with observed galaxy properties. If we identify galaxies in a 
one-to-one mapping with peaks in the density field, our results would suggest that
the orientation of the deformation tensor for galactic-scale perturbations is
correlated with the surrounding mass on scale of the parent cluster. Furthermore, 
the observational study by Vogeley et al. (1992) and Einasto
et al. (1993) have estimated the shape of the fluctuations spectrum on scales of
$\sim 20-100$Mpc, finding a spectral index $n\approx -1.5\pm 0.5$. Which brings as
a possible consequence that the degree of correlation observed in
Figure 2, for flat spectra, suggests that a
correlation of clusters with larger scales should be present. 


    To analise whether this correlation is applyable to the orientation of 
peaks, we calculated the relative orientation between the main axes of the 
deformation tensor and those of the inertia tensor for $R^{(1)}_{f}= 0.6$ 
Mpc. Figure 3 displays the distribution of relative orientation as a function 
of the spectral index. These distributions show that there 
is an important fraction of the peaks whose tensors are similarly oriented; $\sim 
30-50\%$ of the whole sample for $n=-1$ and $n=-2$, and approximately $20-30\%$ 
for $n=1$ and $n=0$. The misalignment of the main axes of the two tensor, would 
have as a consequence that the final morphology of the structures formed from the 
collapse of perturbations, will be poorly correlated with the initial shape of 
their progenitors. It is partially due to this misalignment that the 
perturbations will acquire angular momentum with no correlation between 
neighboring galaxies. 
Because the magnitude of the changes in the 
orientation of the deformation tensor are related to the distribution of mass on 
larger-scales, the degree of misalignment with the inertia tensor increases. So
can hapen with the angular momentum acquisition. Thus, the tidal shear effects 
produced by the large-scale density field will affect the evolution of peaks 
depending on how they are initially oriented with the tidal field.

\section{Effect produced by the surrounding density field}

So far, we have analysed the changes in the orientation of the principal 
axes of the deformation tensor without paying attention in the
magnitude of the deformation itself. We now assess such magnitudes in order to
estimate the importance of the tidal effects produced on the peaks by the 
surrounding mass distribution. This will tell us from which scales the most 
important contribution to the deformation tensors is being produced. 


\subsection{Changes in the density contrast}

We first estimate the tidal field effects on the deformation tensor 
of peaks through the equation which relates the eigenvalues $\lambda_{i}$ 
of the deformation tensor to the density contrast as
\begin{equation} 
      \sum_{i=1}^{3}\lambda_{i}^{(j)}\propto \delta^{(j)},
\end{equation}
where $j$ denotes the values corresponding to different mass scales. The 
difference of these 
\begin{equation}
   \sum_{i=1}^{3}\lambda_{i}\equiv\sum_{i=1}^{3} (\lambda_{i}^{(j+1)}-\lambda_{i}^{(j)})\propto 
   \delta^{(j+1)}-\delta^{(j)},
\end{equation}
shall provides us a measure of the strength of the effect produced by the 
external field. Figure 4 shows the distribution of peaks as a function of the 
difference of the eigenvalues, $\sum\lambda_{i}$. The high number of peaks
at $\sum\lambda_{i}\approx 0$ indicates a weak influence of the nearest 
mass distribution on the deformation tensor by the fact that the eigenvalues 
remain unperturbed. This general behaviour is observed for all the spectral 
indexes. By comparing the changes in the shape of the initial 
distributions of peaks, as the filtering scale increases, it is observed that 
no important correlation exists in the cases $n=1$ and $n=0$. However, for flat 
spectra, the initial shape of the distribution is slightly modified up to scales 
of $\Delta R_{f}\approx 4-5\approx 4$Mpc. After these scales the effect 
decreases significantly as shown by the high number of peaks distributed around 
$\sum\lambda_{i}\approx0$. This sugests again an influence of the large-scale 
superposition of density waves on the peaks. On scales of clusters of galaxies, 
the values $R_{f}\approx 4-5$ would correspond to a correlation of cluster-scale 
perturbation with their environment on scales of $\approx 30-40$Mpc. 

\subsection{Strenght of the tidal field}

We now provide a measure of the changes produced on the deformation tensor, 
by resorting to a concept widely used in the theory of matrix perturbation: 
the norm of the perturbation matrix (e.g. Stewart and Sun, 1990). 
Roughly, the goal of this theory is to predict, or bound, the changes in 
physical processes described by a matrix when the elements of the matrix 
change. The theory would assess, for example, how far the 'magnitude' of a 
matrix associated
to a physical process --e.g. the deformation tensor ${\bf D}_{ij}^{(1)}$--
will change when the elements of the matrix change as ${\bf D}_{ij}^{(2)}= 
{\bf D}_{ij}^{(1)} + {\bf \Delta}_{ij}$, due to a perturbation, ${\bf 
\Delta}_{ij}$. The prerequisite for answering this question is to make precise 
the term 'magnitude', which is done by defining a norm in the space of matrixes. 
We adopted the Frobenius norm defined as 
\begin{equation}
  {\bf F}\equiv\parallel {\bf \Delta}_{ij}\parallel 
         \equiv\sqrt{ \sum_{i,j}\mid\alpha_{ij}\mid^{2} }
         \equiv\sqrt{ \sum_{i}\lambda_{i}^{2}}.
\end{equation}
When the deformation tensor does not suffer 
important changes the norm will tend to take small values. The lower limit
correspond to ${\bf F}=0$ in which the elements of the deformation tensor
are not affected. 


Figure 5 displays the distribution of peaks with norm F. A clear difference in 
the strength of the tidal field is observed  between the different models. The 
maximum of the distribution of peaks, in the cases $n=-1$ and $n=-2$, shows a 
considerably deviation from zero at large radii of filtering, thus confirming the 
influence of the superposition of the density waves which conform the large-scale 
perturbations, on the local deformation of small-scale perturbations. This is 
also seen from Figure 6a, where we plot the mean strength F as a 
function of $R_{f}$. The norm of the deformation tensor at $R_{f}^{(1)}$ was used 
to normalize the maximum effect to unity. We confirmed our prior results: steep 
spectra show a correlation only with the nearest distribution of mass on scales 
smaller than $R_{f}=2$. Flat spectra show important effects coming from even 
larger-scales. An insight of the decaying rate of the effect on density peaks is 
provided by the difference of F's between different shells, which we bined as 
\begin{equation}
\Delta F_{i}=\mid F(R_{f}=8/i)-F(R_{f}=4/i)\mid, \hspace{1.cm} {\rm for 
\hspace{0.5cm}i=4,2,1}
\end{equation}
showed in Figure 6b. It is observed a general tendency for the influence to 
decay. However, the rate of dacaying is faster for those models with steep 
spectra. From these results, one can see that there is no a divergency due to
the mass increasing in the density field. However, it keeps producing important
effects on small-scale peaks. 

\subsection{Density peaks and their environment~?} 

One can expect that the alignment of peaks to show up if the orientions of the 
peaks are compared with the orientation of their host larger fluctuations.
Thus, we have explored the posibility of parallel alignments by 
calculating the cosine of the angle subtended by the main axes of the inertia 
tensor of galactic-scale peaks and the major axes of the inertia tensor 
of the larger-scale fluctuation. 


The small-scale peaks ($R_{f}\equiv1$) used in this analysis were those located 
within a sphere of radius $R_{f}\equiv8$, and whose typical number was $\sim 40$ 
peaks within this region. 10 cases were analised for each of our 6 realizations
of a spectrum with $n=-2$. Some of the results are presented in Figure 7.
A weak parallel alignment was detected in six cases as those marked as [3] 
and [6]. Each of these include $\approx 28\%$ of the total number of peaks. 
However, we could not get any clear insight about the ultimate reason of 
their alignment (e.g. highest peaks, ellipticity, number of neighbours), as 
for allowing us to reproduce them with a major statistical weight in a larger 
realization. What is noteworthy is the existence of some cases for which 
alignment exist.

The even weaker trend for 
parallel alignments showed by the other cases --which include at most 
$10\%$ of the peaks-- is probably not too surprising since we carried out the 
comparison of the orientations with peaks on a scale larger than that where 
we found evidence of a stronger correlation, $\Delta R_{f}\approx4-5$. These 
results therefore, are useful as indicators of an upper limit at which we 
could expect primordial alignments effects. If one adopts a cluster-scale 
smoothing of the density field, the present results would suggest us that 
we cannot expect alignment of the cluster-cluster type on scales larger 
than $40-60h^{-1}$Mpc, in agreement with some observational results: an 
alignment of the main axes of clusters of galaxies has been detected up to 
scales of $30-60h^{-1}$Mpc (e.g. Binggeli 1982; Oort 1984; Plionis 1994) and 
even at the central regions of clusters between the cD galaxy and their host 
cluster (Rhee et al., 1990; West, 1994).

For the steep spectra, the distributions showed no signals for any kind of 
alignment. The density peaks orientations are uniformly distributed.

We have also found that the inertia tensor of about $10-15\%$ of the density 
peaks is oriented with the main axes of the large-scale perturbations on scales 
smaller than $\Delta R_{f}\approx 8$. Then, one should expect that at the 
moment of collapse the inner regions would already be aligned with the host 
structure. The primordial tidal field acting on these perturbations will not 
significantly change their initial orientations because the quadrupole term 
of the peaks, which is the first in getting coupled with the tidal field, would 
be already oriented in that direction.

\section{Discussion}

We have explored the posibility of the existence of primordial alignment effects
which could survive the non-linear evolution of the density field. Some evidence
was detected that such alignments do exist and can be the progenitors of the 
coherence in the orientation of galaxies in clusters and between clusters. The
scales we found, to which the large-scale density fluctuations affect the 
small-scale peaks, nearly match those to which alignment of structures have 
been reported both from observational studies (e.g. Binggeli, 
1982; Oort, 1984; Plionis, 1994; LPG 1987; Muriel and Lambas, 1992) and from 
numerical simulations (e.g. Dekel, West and 
Aarseth, 1984; Barnes and Efstahiou,1987; West, 1989; Little, Weimberg and 
Park, 1991). These scales are $\approx 3-5h^{-1}$Mpc for alignment of 
galaxies and $\approx 30-50h^{-1}$Mpc for clusters of galaxies. 

The weak evidence of alignments we found resulted valid only for flat spectra,
and therefore a observational confirmation of alignments to significant 
statistical levels could be used as an evidence of the flatness of the spectrum 
at the scales observed, but it cannot determine the exact value of the spectral 
index.

A natural consequence for the significant fraction of peaks with a misalignment 
between the inertia and the deformation tensors, is that they are expected not 
only to acquire angular momentum but also to suffer important morphological 
changes depending on the strength of the tidal field. A question which arise 
from the misalignment is wheter it is the shape of the peaks or it is the 
deformation tensor which determine the way in which the peak collapse. If the 
latter is the dominant mechanism, perturbations will be able to acquire angular 
momentum from the large scale density field, without being correlated with 
their nearest neighbours. 

A clue on the origin of correlations to different scales was
given by Bond (1986, 1987). He showed that for Gaussian
random fields, density peaks on scale of clusters have triaxial tails which
can extend to $\sim 20h^{-1}$Mpc. The nonlinear evolution of these density peaks
would enhance that initial anisotropy producing that the clusters tend to be 
born aligned  with their surroundings (West, Dekel and Oemler, 1989; West, 
Villumsen and Dekel, 1991; Barnes, 1992). This probably is the origin of the alignments of
neighbouring clusters. An attempt for following the 
origin and evolution of the alignment of structures is being carried out by Gonz\'{a}lez 
and Col\'{i}n (1997).

\clearpage

\figcaption[fig1.ps]
{Examples of the changes in the orientation of the main axis of the 
deformation tensor at the density peaks positions.} 

\figcaption[fig2.ps]{{\it (a)} Histograms of the changes in the orientation 
of the deformation tensor, as a function of the shells distance, for
$n=1,0$, {\it (b)} for $n=-1$ and {\it (c)} for $n=-2$.}

\figcaption[fig3.ps]{Histograms showing the misalignment between the main axes 
of the deformation tensor and those of the inertia tensor.}

\figcaption[fig4.ps]{Histograms of 
$\sum_{i=1}^{3}\lambda_{i}=(\lambda_{i}^{(j+1)}-
\lambda_{i}^{(j)})$ as a function of the shell distance. The cases $n=-1,-2$
show important changes at the distances indicated in the panels.}

\figcaption[fig5.ps]{Histograms of the Frobenius norm, as a function of the
shell distance and the spectral index. For $n=0$, $F=0$ suggest no important
changes coming from the scales indicated in the panels.}

\figcaption[fig6.ps]{{\it (a)} Decaying rate of the Frobenius norm as 
a function of the shell distances. {\it (b)} Decaying rate of F when the shell
distances are binned as indicated in the text.}

\figcaption[fig7.ps]{``{\it Clusters of peaks}'' showing the orientational
distribution of their members. Cases [3] and [6] show a week tendecy for
alignment.}
\clearpage 

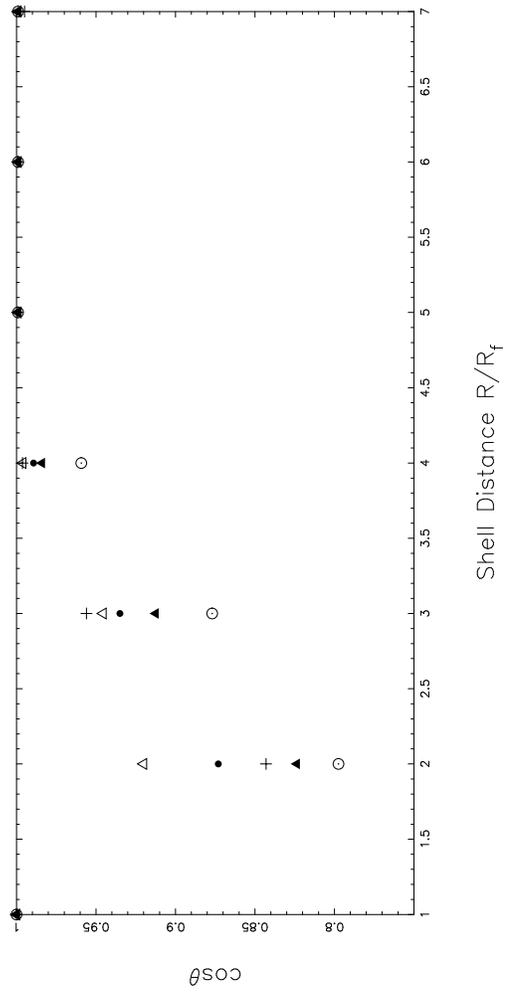
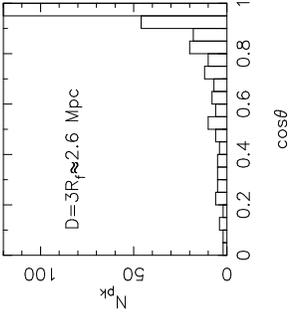
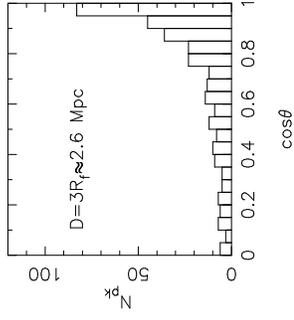
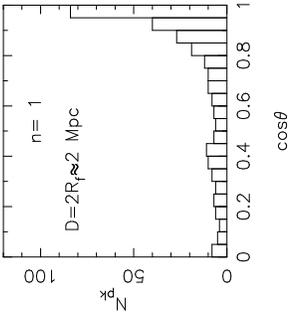
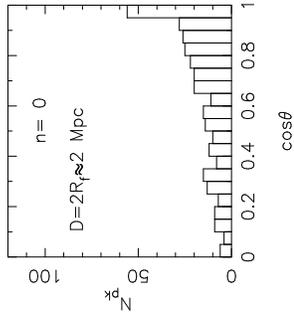
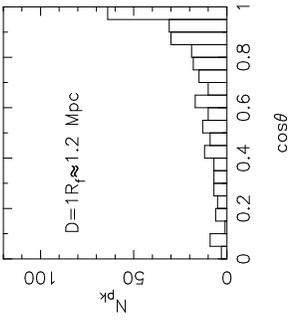
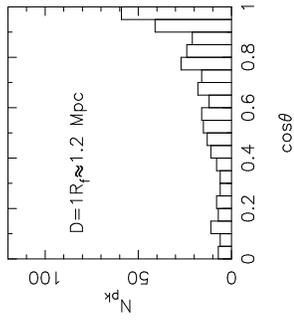
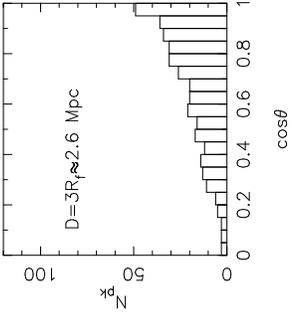
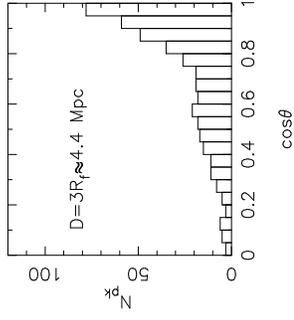
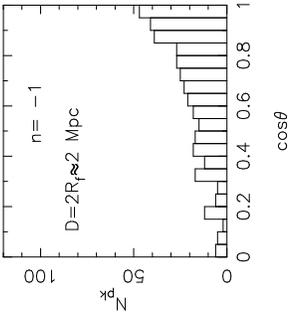
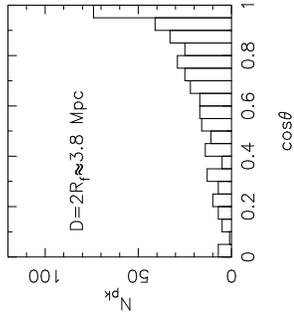
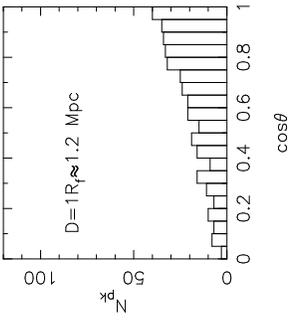
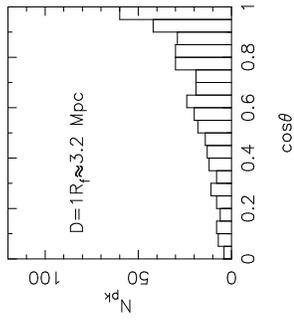
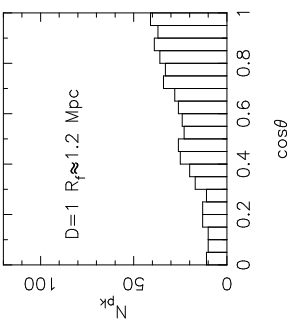
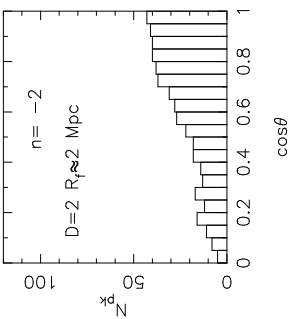
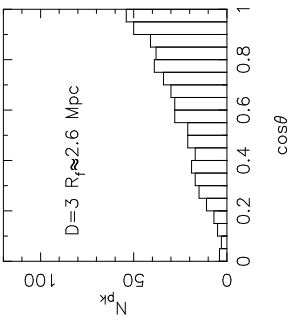
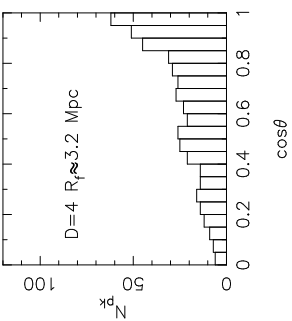
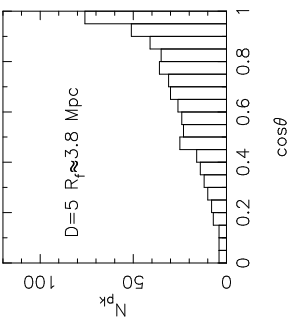
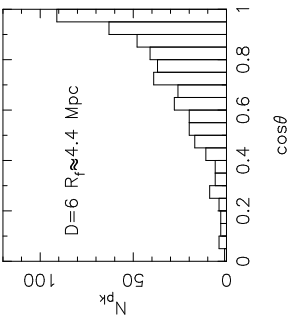
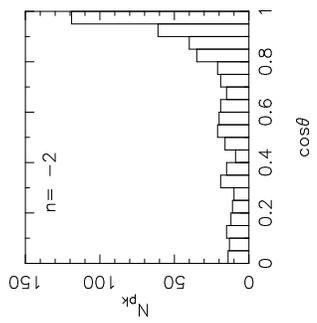
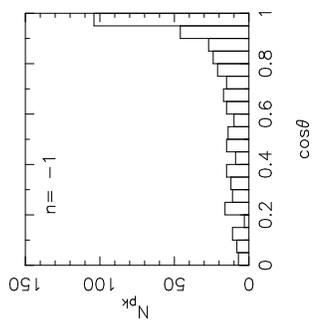
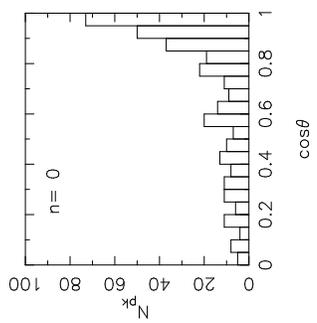
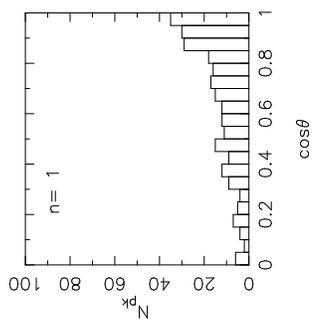
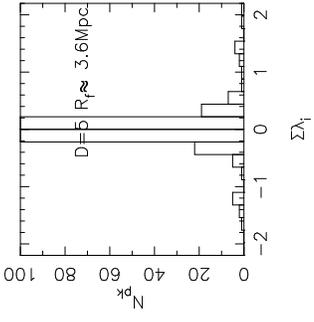
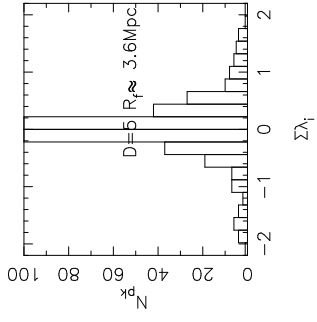
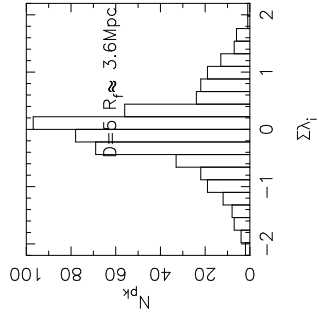
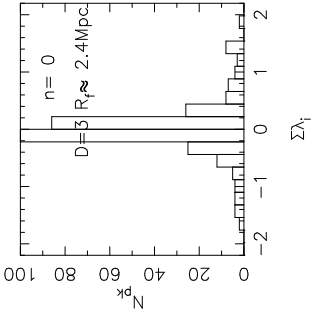
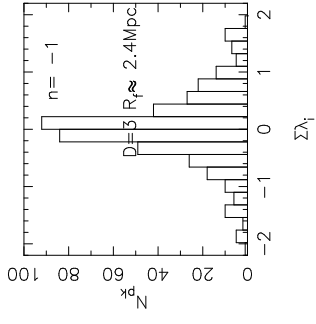
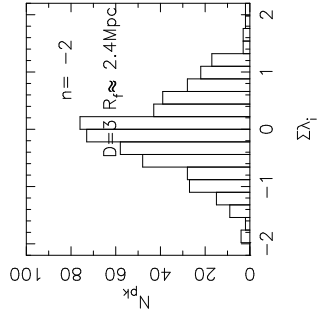
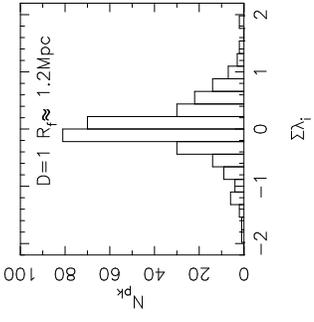
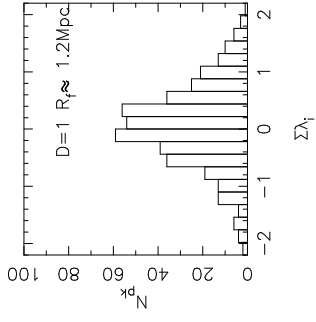
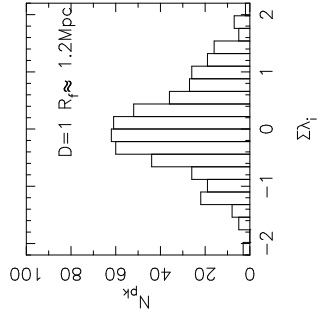
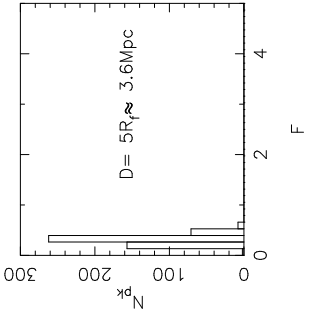
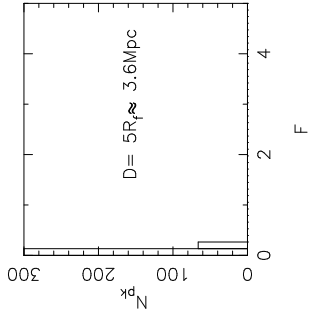
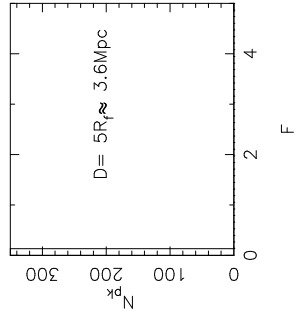
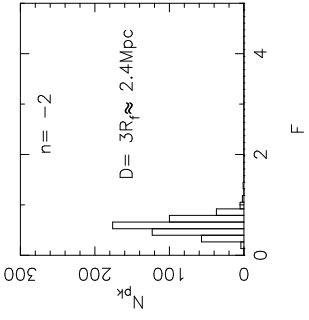
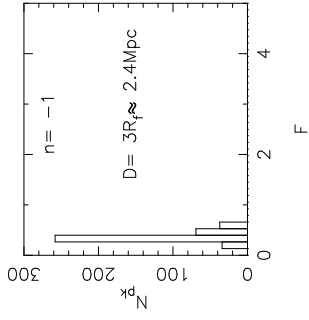
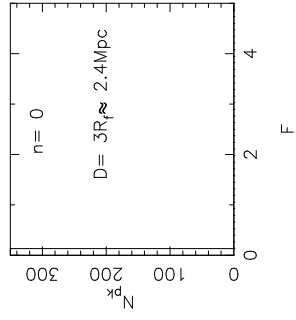
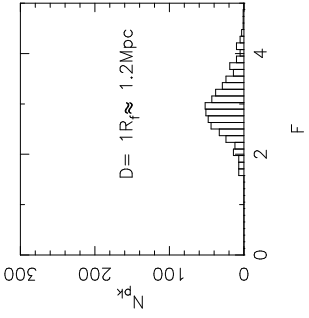
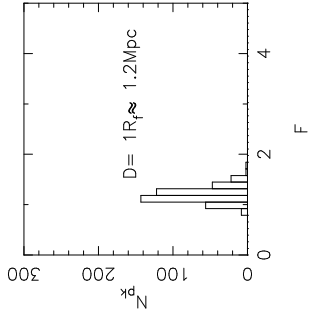
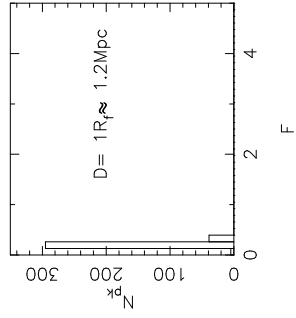
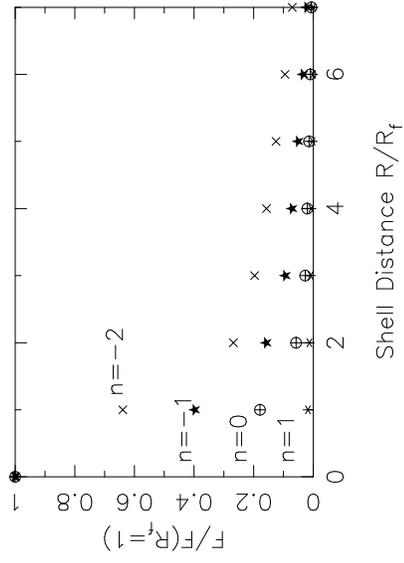
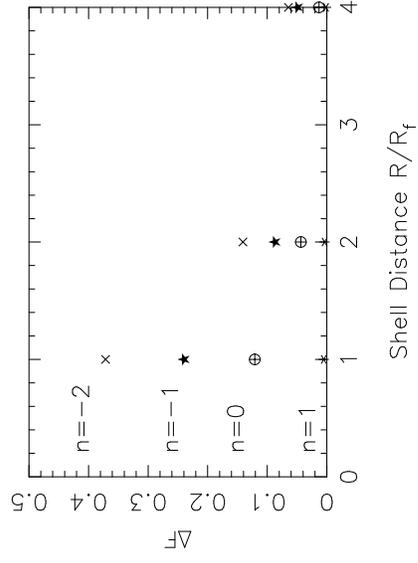
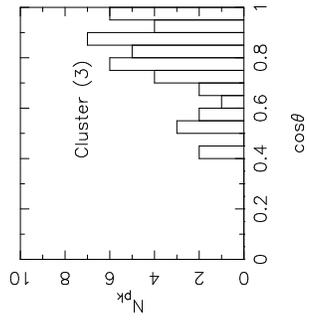
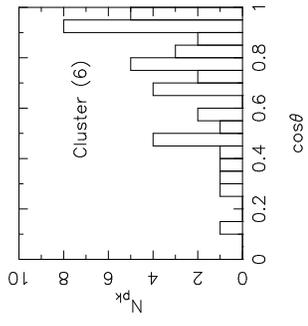
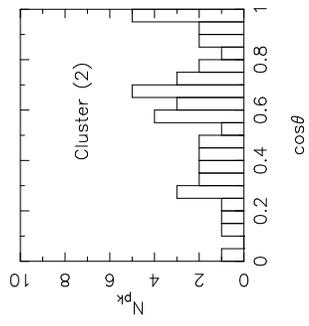
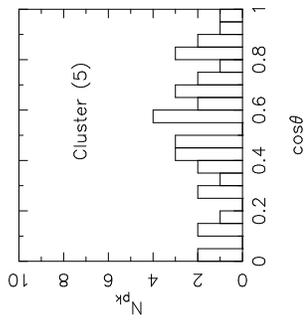
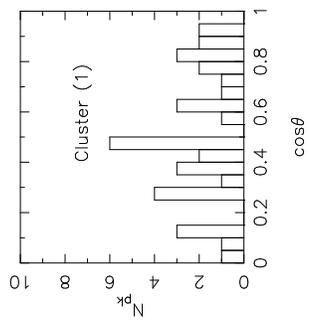
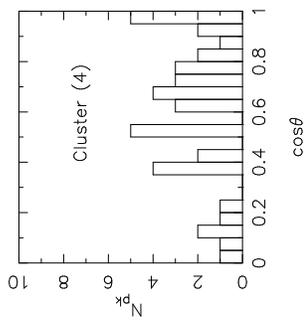
\end{document}